\documentclass[conference]{IEEEtran}
\usepackage{graphicx,cite}
\usepackage{epsfig}
\usepackage[cmex10]{amsmath}
\usepackage{amssymb}
\usepackage{array}
\usepackage{fixltx2e}
\usepackage{amsfonts}
\usepackage{lineno}

\usepackage{marginnote}

\begin{document}
\title{A Wireless Channel Sounding System for Rapid Propagation Measurements}

\author{\IEEEauthorblockN{Muhammad Nazmul Islam}
\IEEEauthorblockA{Department of Electrical \& Computer Engineering\\
WINLAB, Rutgers University, NJ, USA \\
Email: mnislam@winlab.rutgers.edu}
\and
\IEEEauthorblockN{Byoung-Jo J. Kim, Paul Henry, Eric Rozner}
\IEEEauthorblockA{AT\&T Labs Research \\
Middletown, NJ, USA\\
Email: \{macsbug,psh,erozner\}@research.att.com}}

\maketitle

%
%

\begin{abstract}

Wireless systems are getting deployed in many new environments 
with different antenna heights, frequency bands and multipath conditions. 
This has led to an increasing demand for more channel measurements to
understand wireless propagation in specific environments and assist deployment engineering. 
We design and implement a rapid wireless channel sounding system,
using the Universal Software Radio Peripheral (USRP) and GNU Radio software,
to address these demands. Our design measures channel propagation
characteristics simultaneously from multiple transmitter locations.
The system consists of multiple battery-powered transmitters and receivers.
Therefore, we can set-up the channel sounder rapidly at a field location
and measure expeditiously by analyzing different transmitters' signals during 
a single walk or drive through the environment.
Our design can be used for both indoor and outdoor 
channel measurements in the frequency range of $1$ MHz 
to $6$ GHz.
We expect that the proposed approach, with a few further refinements, 
can transform the task of propagation measurement as a routine part of 
day-to-day wireless network engineering\footnote{(c) 2012 AT\&T Intellectual Property. All rights reserved.}.

\end{abstract}

\section{Introduction}

Radio propagation measurement is essential for developing propagation models~\cite{Cox} and crucial for 
deploying and maintaining operational systems. There are numerous path loss and delay spread models based on measurements made in a variety of environments, using different transmitter heights and carrier frequencies~\cite{Hata,Herring,ITM,Cost231,Rappaport,Sridhara}. However, most models have substantial errors when compared to reality~\cite{Phillips}, due to the complicated nature of wireless propagation through complex environments. Besides, a new environment to deploy wireless systems (e.g., small-cell systems with low base stations in urban areas), often requires a measurement campaign to determine if existing models can be adapted to it, and if so, what parameters must be changed. 

In deployment of wireless systems in complex, high-value locations (such as dense urban areas, or unique and complex buildings), propagation modelling or even ray-tracing prediction tools may not account for important and unpredictable realities of target locations. At least spot verifications of the modelling outputs and calibrations, by means of a small measurement campaign, may be required. In addition, sophisticated modeling approaches~\cite{Rappaport,Sridhara} may require 3D models of the environment or the associated material properties, which may be difficult and expensive to obtain and yet may not account for hidden metal structures or openings.

In short, it would be ideal if wireless propagation measurements were so simple, inexpensive and quick that measuring an environment would often be the first choice by radio engineers, rather than modeling and verifying. Short of that ideal goal, we argue that the current wireless channel measurement practices can be made far simpler and quicker so that more measurements can be taken at modest cost to support further modelling, design and deployment efforts. We would like to achieve this without sacrificing too much in accuracy or scope of measured parameters, compared to the conventional channel measurement campaigns using laboratory equipment or custom-built apparatus.

In this paper, we describe our first effort in designing a rapid wireless channel sounding system.
We also mention our rapid prototyping exercise that was enabled by the recent
advancements of software-defined radio (SDR) technologies. We use Universal Software Radio Peripheral (USRP)~\cite{USRP} and GNU Radio~\cite{GNUradio} as the SDR platform. We perform
sliding correlator and frequency domain channel sounding in indoor and outdoor
experiments respectively.



\section{A CASE FOR RAPID CHANNEL SOUNDING SYSTEM}

\subsection{Limits of propagation models}

Most easy-to-use channel models~\cite{Hata,Herring,ITM,Cost231} are statistical 
in nature, and thus almost certain to have errors at specific locations. 
More location-specific modelling approaches such as ray-tracing~\cite{Sridhara} or partition-based models~\cite{Rappaport} require more detailed information on the environment. In all cases, at least some measurements are often required for calibration and verification. Some complex deployment environments might benefit from more extensive measurements, if such measurements can be made rapidly and cheaply by field personnel without extensive training.

\subsection{Benefits of using multiple transmitters}

An important decision during deployment engineering would be where to place base stations in a given environment, given coverage and capacity requirements. Using multiple transmitters at many candidate locations emitting reference signals that can be distinguished by the receiver, a single walk or drive through the environment enables rapid measurements taken for all candidate locations. Also, if the effect of different antenna heights, carrier frequencies, antenna types or mounting arrangements need to be measured, multiple transmitters can be set up with different parameters of interest to quickly measure during a single run through the areas of interest.

\subsection{Essential features}

\subsubsection{Reference signals}

The reference signals from multiple transmitters should be distinguishable even when their respective received power levels differ by large amounts at the receiver, as the receiver can be close to one transmitter while measuring a far-away transmitter at the same time. 

\subsubsection{Battery-powered operation}

Arranging power supply to many transmitter locations before starting measurements can be burdensome in most environments. Thus, the fixed transmitters must be battery-powered for several hours with a reasonably-sized battery.

\subsubsection{Small form factor}

To further ease the preparation and measurement, the transmitters and the receivers should be reasonably small and light so that they can be quickly and safely placed or mounted on desired locations.

\subsubsection{Low cost}

Having to use multiple transmitters and receivers, it would be desirable to keep the cost of individual units low. 

\subsubsection{Flexibility}

The approach should be flexible to support channel measurement across
a wide frequency region.


\subsection{Our contributions}

Our approach incorporates the simultaneous measurement of channel propagation
characteristics from multiple transmitter locations. 
Previous works in the related literature~\cite{Cox,Jana,Howard,Porrat,Rappaport,Firooz}
focused on channel sounding measurements from a single transmitter location. 
To the best of our knowledge, ours is the only work in the literature
that implements multiple transmitter simultaneous channel sounding.

Traditional channel sounding systems use expensive
measurement equipment like
vector network analyzers~\cite{Jana,Howard},
vector signal generators~\cite{Phillips},
spread spectrum channel sounders~\cite{Rappaport}, etc. 
Due to the open source GNU Radio software and inexpensive USRP radios,
our proposed channel sounding system costs significantly less than these systems.

The carrier frequency of the USRP daughterboards can be
varied from $1$ megahertz (MHz) to $6$ gigahertz (GHz). Therefore, our approach
can perform channel measurements in this wide frequency range.

\begin{figure}[t]
\centering
\includegraphics[scale=0.35]{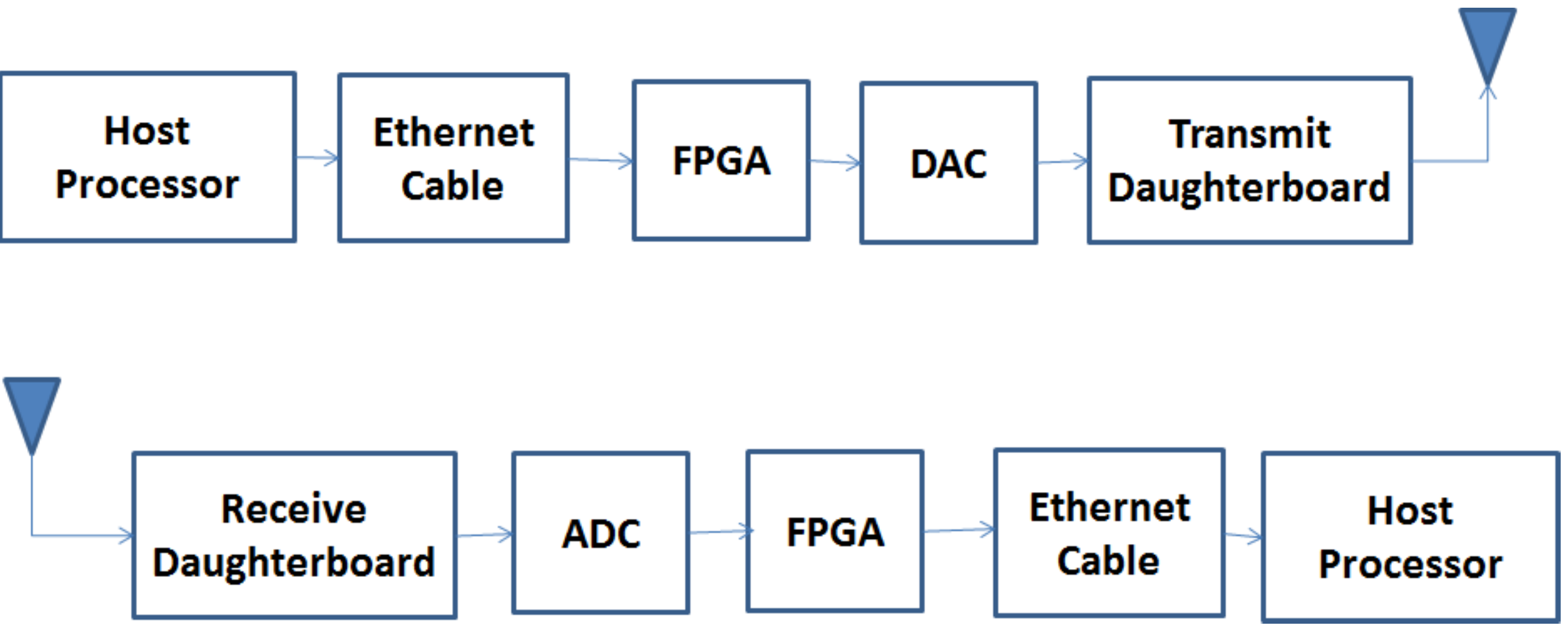}
\caption{USRP block diagram}
\centering
\label{fig:USRP_block_diagram}
\end{figure}

\section{Measurement System}


We use GNU Radio software and USRP daughterboards in the
channel sounding experiments. The top and bottom parts of Fig.~\ref{fig:USRP_block_diagram}
show the transmit and receive block diagrams of the USRP respectively.
In the transmitter side, the host processor sends complex baseband samples
to the field programmable gate array (FPGA) through an ethernet cable. The FPGA board low pass
filters and up-converts the signal to a higher sampling rate. 
Thereafter, the signal goes through the digital-to-analog converter (DAC)
and passband frequency conversion stage to the transmitter antenna. 
The receiver side operates exactly in the opposite manner. 

The host processor views these complex baseband samples as floating
point numbers. GNU Radio is an open source software that allows the use of
digital signal processing algorithms on these floating point numbers.

We use USRP networked (N) and embedded (E) series radios for our indoor 
and outdoor experiments respectively. 
The N series requires an external laptop for operation. On the other hand, 
the E series radios contain an embedded processor and 
can work as stand-alone pre-programmed transceivers. 
USRP radios can offer up to 400 megasample per second (MS/s) and 100 MS/s
sampling rate in the DAC and analog-to-digital converter (ADC) respectively.
However, the speed bottleneck of the GNU Radio software
usually comes from the host processor speed and the capacity of the ethernet cable.
The N series and E series software allow up to 50 MS/s and 8 MS/s 
data transfer rate from the host processor respectively. 
The N and E series are used in sliding correlator and frequency domain
channel sounding systems respectively.



\section{Sliding Correlator Channel Sounding}

\subsection{Methodology}

\begin{figure}[t]
\centering
\includegraphics[scale=0.30]{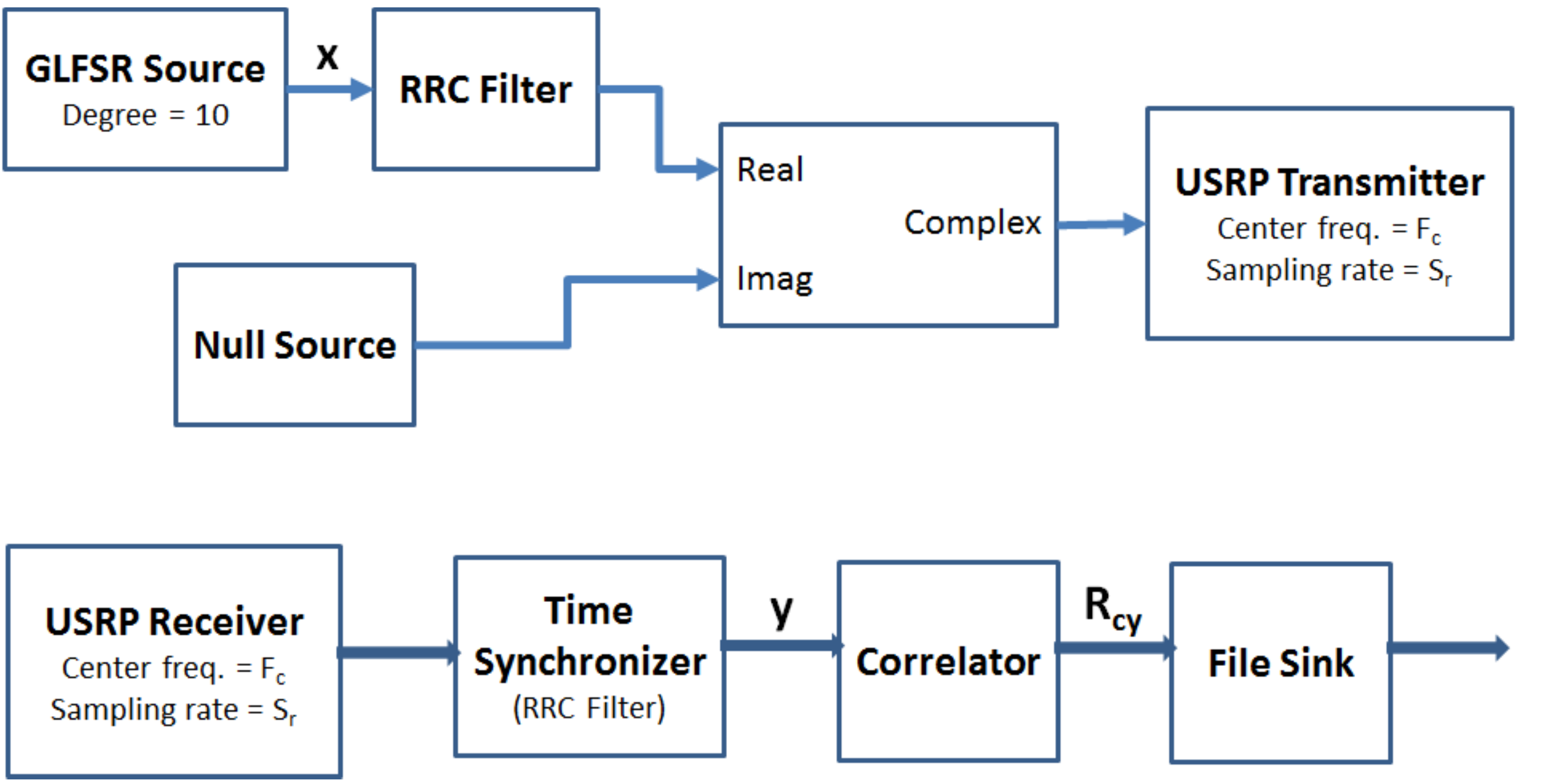}
\caption{Sliding Correlator Channel Sounder System}
\centering
\label{fig:TimeDomainSystem}
\end{figure}

In the sliding correlator approach, the transmitters send a pseudo-noise (PN)
sequence with a $60$ nanosecond (ns) pulse duration and the receiver obtains the wideband path loss
and multipath delay profile. 
The upper and lower parts of Fig.~\ref{fig:TimeDomainSystem} 
uses a single transmitter and receiver to show the 
transmit and receive diagrams of the sliding correlator system respectively . 

\subsubsection{Transmission}

The transmitter sends $\mathbf{x}$, a Galois linear feedback shift register (GLFSR)
maximal length PN sequence of degree 10. $\mathbf{x}$ can be analytically written as follows:
\begin{equation}
x[n] = \sum_r c[n-rN]      \label{eq:x[n]}
\end{equation}
where, $N = 1023$ and $\mathbf{c}$ is a chip sequence of length 1023,
$\mathbf{c} = [c_0,\cdots,c_{1022}]$.
%
Here, $c_i \, \forall \, i \, \in \, [0,1022]$ takes the value of either $+1$ or $-1$.
Defining $\mathbf{R}_{cx}$ as the correlation output of $\mathbf{c}$
and $\mathbf{x}$ and using the properties of
PN sequence, 
\begin{equation}
R_{cx}[n] = \left\{
\begin{array}{c l}      
    1 & n = 0, N, -N, 2N, -2N, \cdots \\
    -\frac{1}{N} &  otherwise
\end{array}\right\}    \label{eq:R_cx}
\end{equation}
The signal $\mathbf{x}$ is passed through a root raised cosine (RRC) filter
and then sent to the real input of the USRP transmitter module. 
The imaginary input comes from a null source. The USRP transmitter module
sends the complex baseband samples to the USRP transmit path 
and establishes the transmit frequency and sampling rate.
The baseband equivalent transmitted signal is given by:
\begin{equation}
x(t) = \sum_n x[n] p(t-nT_s)   \label{eq:x(t)}
\end{equation}
where $T_s$ is the period of the RRC generated pulse.

\subsubsection{Multipath Channel}

The impulse response
of the multipath channel can be written as:

\begin{equation}
h(t) = \sum_{l=0}^{L-1} \alpha_l \delta (t - \tau_l)    \label{eq:h(t)}
\end{equation}

Here, $L$ is the number of multipaths in the channel. 
$\alpha_l$ is the complex gain of each multipath and $\tau_l$ is the associated 
delay. We assume $\tau_0 = 0$ since we focus on relative delay.

\begin{table}
\begin{center}
\begin{tabular}{|l|l|} \hline
Parameter & Value \\ \hline
Pathloss dynamic range & 45-105 dB \\ \hline
Temporal resolution & 60 ns \\ \hline
Maximum multipath delay & 61 millisecond \\ \hline
\end{tabular}
\end{center}
\caption{(Sliding Correlator Channel Sounder Design Parameters)} \label{tab:SlidingCorrelator}
\end{table}

\subsubsection{Reception}

The baseband equivalent received signal, in time domain, is obtained by:
\begin{equation}
y(t) = (x *  h)(t) = \sum_{l=0}^{L-1} \alpha_l x (t - \tau_l) \label{eq:y(t)}
\end{equation}
%
$y(t)$ goes through the USRP receive path, gets sampled and arrives
at the USRP receiver module of Fig.~\ref{fig:TimeDomainSystem}.
The time synchronizer block finds the proper phase of 
the RRC pulses.
A detailed theoretical description of the time synchronizer 
can be found in~\cite{Harris}
and the open source code description can be found in~\cite{Polyphase}.
Proper timing synchronization leads to the following discrete
received signal $\mathbf{y}$,
\begin{equation}
y[j] = y[jT_s] = \sum_{l=0}^{L-1} \alpha_l \sum_n x[n] p[(j-n)T_s - \tau_l]  \label{eq:received1}
\end{equation}
%
%
Equation~\eqref{eq:received1} follows from~\eqref{eq:x(t)} and~\eqref{eq:y(t)}. 
Assume that the multipath delay $\tau_l$ is an integer multiple 
of the pulse period $T_s$.
With this assumption, $\tau_l = c_l T_s$ where $c_l$ is a non-negative integer. 
The properties of the RRC filter suggest that $p(nT_s) = 0$ if $n \neq 0$~\cite{Proakis}.
Therefore,
\begin{eqnarray}
y[j] & = & \sum_{l=0}^{L-1} \alpha_l \sum_n x[n] p[(j-n-c_l)T_s] \nonumber  \\
& = & \sum_{l=0}^{L-1} \alpha_l x[j-c_l]  \label{eq:received2}
\end{eqnarray}
Now, the correlator block produces, 
%
\begin{equation}
R_{cy} [n] = corr(\mathbf{c},\mathbf{y}) = \sum_{l=0}^{L-1} \alpha_l R_{cx} [n - c_l]  \label{eq:R_cy}
\end{equation}
where, $\mathbf{R}_{cx} = corr(\mathbf{c},\mathbf{x})$.
%
%
Using \eqref{eq:R_cx} in \eqref{eq:R_cy}, one
can easily find the complex multipath gain $\alpha_l$
at delay, $\tau_l = c_l T_s$. For example, $R_{cy} [0], R_{cy} [N], \cdots$
lead to the calculation of $\alpha_0$ whereas, $R_{cy}[1], R_{cy}[N+1], \cdots$ lead
to $\alpha_1$. The multipath power-delay profile can be obtained from the powers of the individual multipath components $|\alpha_l|^2$.
The path loss can be found from the difference of the known transmit power and the total power 
in the multipath components ($\sum_{l=0}^{L-1} |\alpha_l|^2$). 

Our approach probes the channel only at the
integer multiples of the pulse duration, $T_s$. However, the multipath
rays can arrive at other times, as well. Therefore,
our observation at time $T_s, 2 * T_s, \cdots$
is actually an effect of multipath at nearby times. 
The author of~\cite{Jean} theoretically estimates the number of
arrival paths and the associated attenuation and delay from
the channel gain at the integer multiples of the pulse duration.
The implementation of~\cite{Jean} in GNU Radio USRP framework
remains an area of future research.

\begin{figure}[t]
\centering
\includegraphics[scale=0.35]{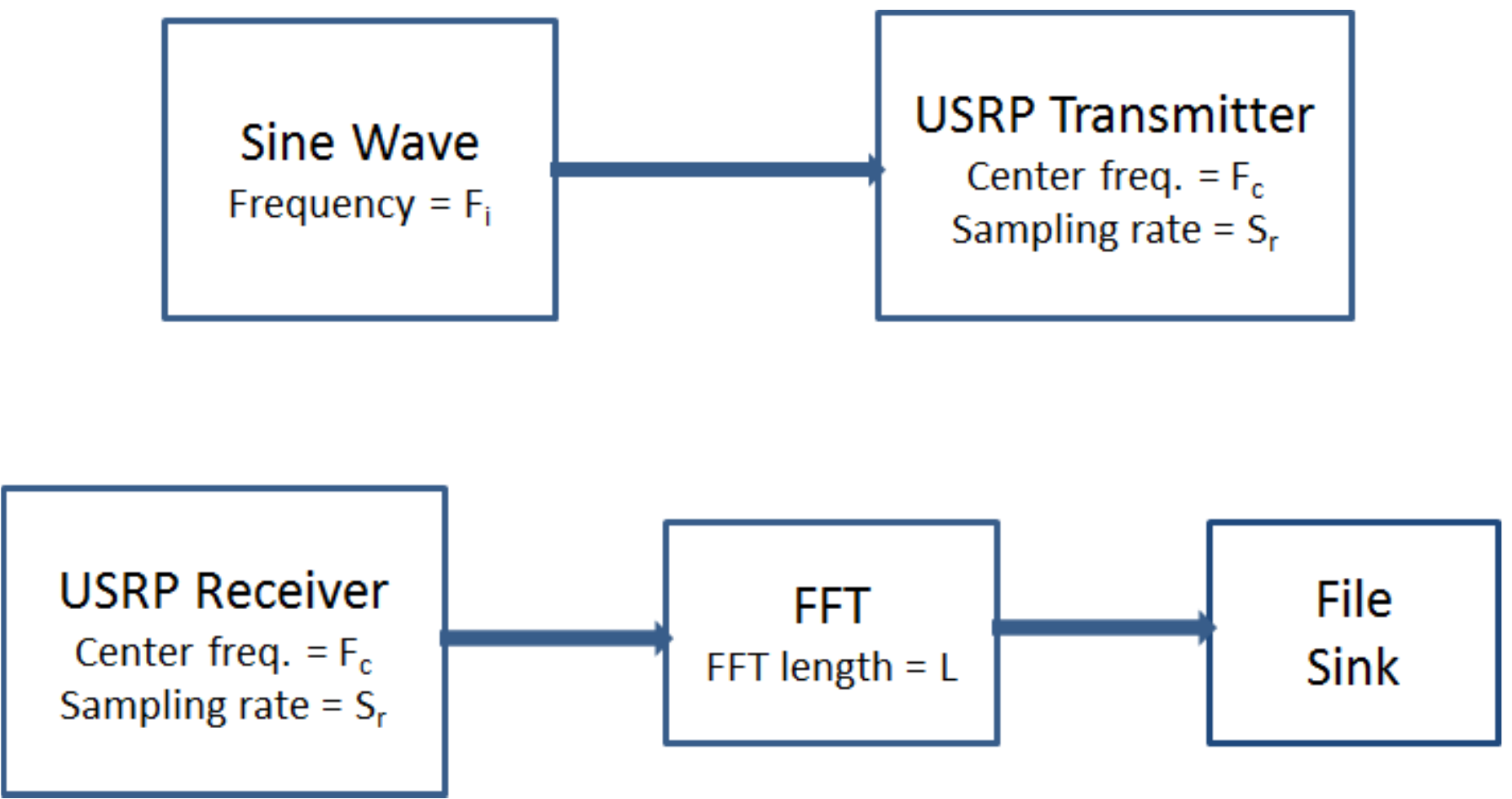}
\caption{Frequency Domain Channel Sounder System}
\centering
\label{fig:FrequencyDomainSystem}
\end{figure}

\begin{table}
\begin{center}
\begin{tabular}{|l|l|} \hline
Parameter & Value \\ \hline
Pathloss dynamic range & 45-120 dB \\ \hline
Frequency separation $(\Delta f)$ & 2 MHz \\ \hline
Number of steps $(N)$ & 10 \\ \hline
Temporal resolution & 27.8 ns \\ \hline
\end{tabular}
\end{center}
\caption{(Frequency Domain Channel Sounder Design Parameters)} \label{tab:FreqParameters}
\end{table}

\subsection{Multiple transmitter sliding correlator channel sounding algorithm}

Different transmitters repeatedly access their allotted time 
slots and transmit the GLFSR PN sequence. The receiver captures the PN sequences from each 
transmitter and finds the path loss and delay profile for each of them.
The overall algorithm is summarized below:

\begin{enumerate}


\item Assume there are $N$ transmitters. Transmitter $i$
transmits in the desired frequency band during the time slot $[t_{i-1}+r*T_p \, 
, \, t_i+r*T_p] \, \forall \, r \, \in \, [0,1,\cdots,M]$.
Here, $t_i - t_{i-1} = \Delta t$ is the allotted time slot length of
each transmitter during each time period $T_p$ and $T_p = \Delta t \times N$.
Also, $M \times T_p$ is the total experiment duration.

\item The user opens the floor map image in the receiver laptop and
clicks a point that corresponds to the present location.
The receiver flow graph is initiated at time $p \times T_p$
where $p$ is the nearest integer. 

\item The receiver captures the samples during the time slot $[p T_p, (p+1) T_p]$,
splits the floating point numbers into $N$ segments
and uses the $i^{th}$ segment to calculate the
path loss and delay profile of the $i^{th}$ transmitter.

\item The wideband path loss, delay profile and X \& Y
coordinates of the location are stored in the laptop.


\end{enumerate}

\begin{figure}[t]
\centering
\includegraphics[scale=0.6]{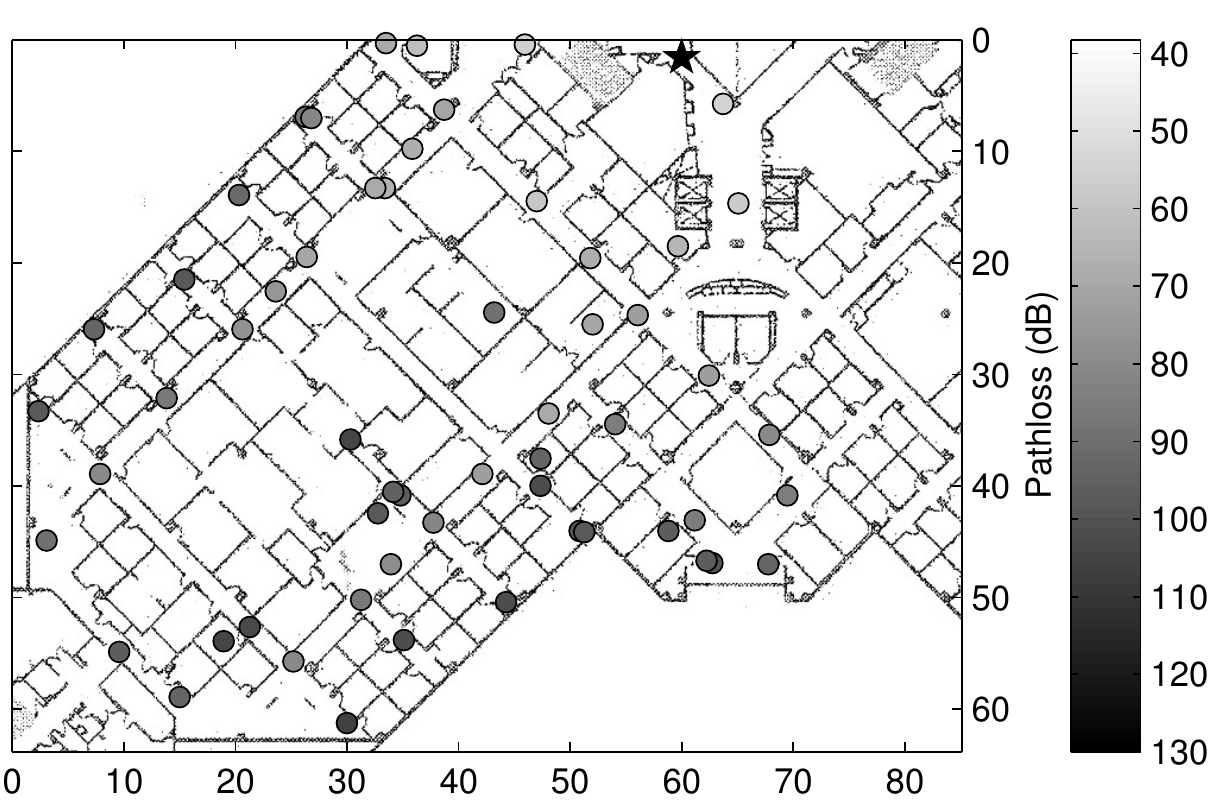}
\caption{Path loss data for indoor transmitter 1}
\centering
\label{fig:CenterTxData}
\end{figure}

\subsection{Challenges of multiple transmitter channel sounding in the sliding correlator method}

\subsubsection{Time synchronization}

We use the USRP N series radios, controlled by external laptops, 
in the sliding correlator channel sounding method.
The clock timing of these laptops is synchronized in advance through 
network time protocol (NTP) servers~\cite{NTP}. The synchronized
laptops control the TDMA operation of the multiple transmitters.

\subsubsection{Near-far effect}

In an ideal $N$ transmitter TDMA system, $N-1$ transmitters remain silent when
one transmits. In order to implement this method in our setup,
the $N-1$ USRP radios have to either turn off or transmit null source
during the active transmission period of the other radio. 
The frequent turn on-and-off leads to the freezing up of USRP radios.
On the other hand, USRP radios leak a small amount of power while
transmitting a null source. This leakage power leads to the classical
near-far problem in a multiple transmitter scenario, i.e., 
the channel measurements of the far transmitter get
overwhelmed by the leakage from the near transmitter, due to
the large difference of path loss among the transmitters.
In order to avoid these two problems, we take the following approach:
when transmitter $i$ transmits, transmitter $j \, \forall \, \in  \, [1,N] \, , \, j \, \neq \, i$
transmits in the industrial, scientific and medical radio band
at the lowest power possible. 
The receiver receives samples in the desired frequency band
and therefore, the channel measurements of different transmitters
remain independent of each other.

The design parameters of the sliding correlator channel sounder are 
given in Table~\ref{tab:SlidingCorrelator}.

\section{Frequency Domain Channel Sounding}

The USRP N series radio based sliding correlator channel sounding system 
requires an external laptop for each transmitter and receiver. 
The use of an external laptop in outdoor environments
is inconvenient due to its heavy weight and limited battery lifetime.
Therefore, we use USRP E series transmitters in outdoor
experiments. The embedded processor of E series radios
can provide sampling rates up to 8 MS/s. In a sliding correlator channel 
sounding system, this sampling rate limits the temporal resolution
to $250$ ns. This resolution is too low to handle
the rich multipath delay spread of an outdoor environment.
Therefore, we perform frequency domain channel sounding
in outdoor experiments.

\subsection{Methodology}

In the frequency domain channel sounding method, 
the transmitters and the receiver synchronously sweep a given frequency
band in $N$ discrete steps of $\Delta f$. 
By sweeping a large frequency band, one
can obtain a very fine temporal resolution~\cite{Jana}. 
The top and bottom parts of Fig.~\ref{fig:FrequencyDomainSystem} show
the transmitter and receiver block diagrams of a
frequency domain channel sounder using a single transmitter and receiver. 




\subsection{Multiple transmitter frequency domain channel sounding algorithm}

\begin{figure}[t]
\centering
\includegraphics[scale=0.6]{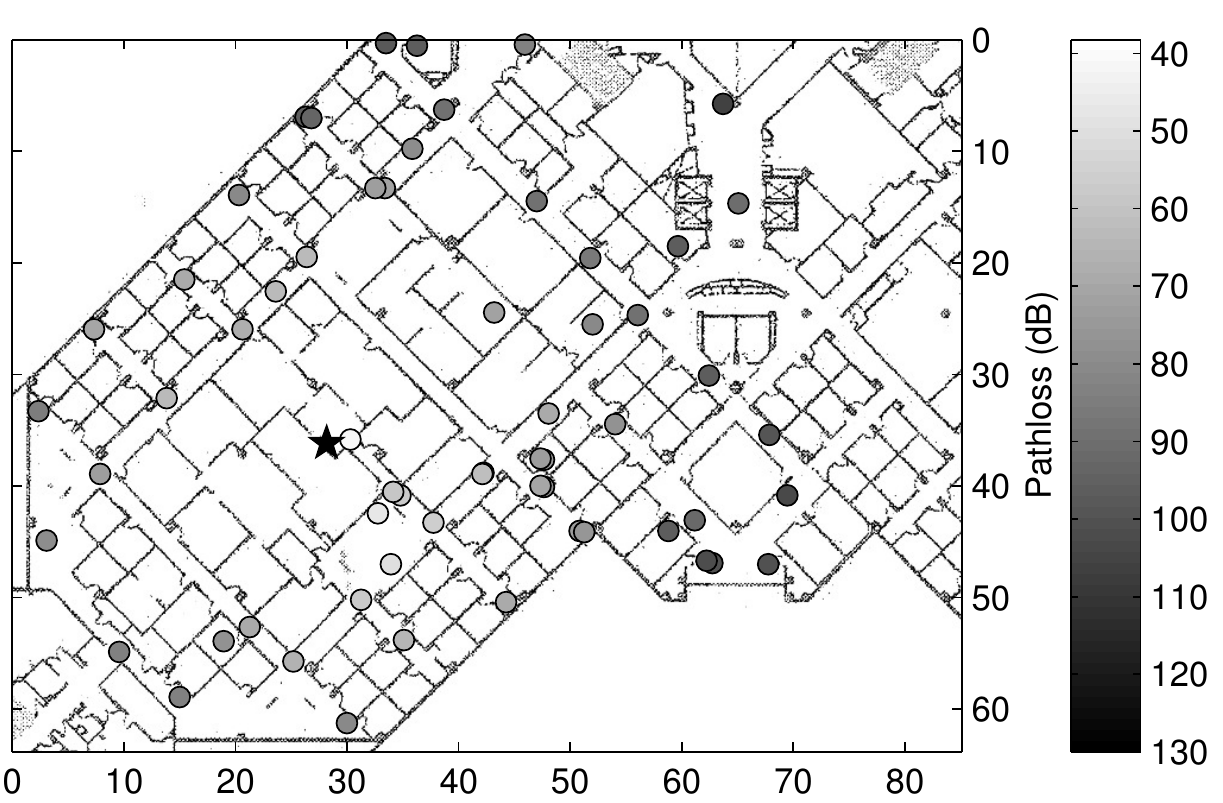}
\caption{Path loss data for indoor transmitter 2}
\centering
\label{fig:PatlabTxData}
\end{figure}

Assume that there are $N$ carrier frequency steps and $K$ transmitters. There is a
predefined list of carrier frequencies, $\mathbf{F}_c = [F_{c_1},\cdots,F_{c_N}]$
and sinusoidal frequencies, $\mathbf{f} = [f_1,\cdots,f_K]$.

\begin{enumerate}

\item The USRP clocks of the transmitters and the receiver get time synchronized
on-the-fly through global positioning system (GPS)~\cite{GPSsync}. 

\item Transmitter $k$ steps through the carrier frequency list 
and transmits a sinusoidal tone at $[F_{c_1}+f_k,\cdots,F_{c_N}+f_k]$
frequencies in $N$ steps.

\item The receiver synchronously steps through the carrier frequency list with sampling rate $S_r$
and performs an FFT of length $L$ on the received samples.

\item Transmitter $k$'s signal falls in the $\frac{L \times f_k}{S_r}$ bin of the FFT. 
 The narrowband path loss of the $k^{th}$ transmitter at frequencies $[F_{c_1}+f_k,\cdots,F_{c_N}+f_k]$
is found based on the power in the corresponding bin.

\end{enumerate}

\subsection{Challenges of multiple transmitter channel sounding in
the frequency domain method}

Theoretically, a large number of complex sine waves can be 
accommodated in the Nyquist transmission band $[-\frac{S_r}{2},+\frac{S_r}{2}]$.
However, some of the power in a tone from a given transmitter
can leak into adjacent frequency regions due to phase
noise and other imperfections. Hence, the transmitters'
sinusoidal tones need to be separated by a guard band
so that the path loss calculations of different transmitters remain independent of each other.
This guard band, along with the maximum sampling rate of the receiver,
limits the maximum number of transmitters to 5-6 in one time frame.
However, separation of the transmitters in
both time and frequency domain, can accommodate a large number
of transmitters in the frequency domain channel sounding method.


\subsection{Mean wideband path loss}

The frequency domain approach provides narrowband path losses in the
frequency range $[F_{c_1}+f_k,\cdots,F_{c_N}+f_k]$. The wideband path
loss in this frequency band can be obtained by taking the average of
the individual path losses.

The design parameters of the frequency domain channel sounder are given in 
Table~\ref{tab:FreqParameters}.

%
%

\section{Experimental Results}

\begin{figure}[t]
\centering
\includegraphics[scale=0.7]{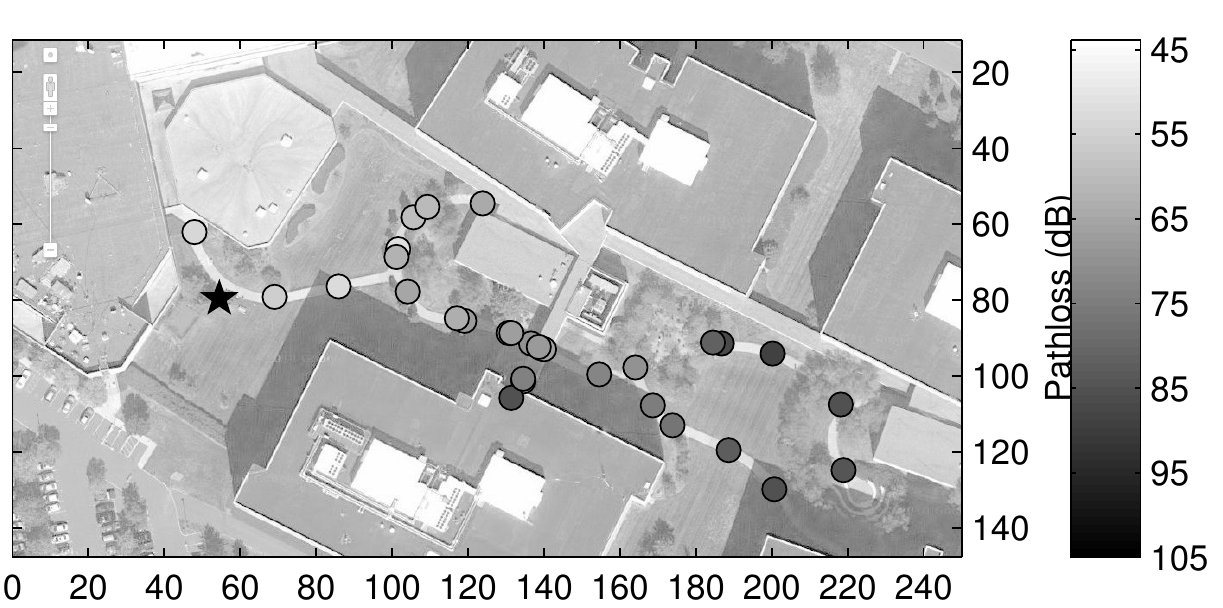}
\caption{Path loss data for outdoor transmitter 1}
\centering
\label{fig:TripodTxData}
\end{figure}

\subsection{Sliding correlator channel sounding results}

The indoor channel measurements were obtained using the sliding
correlator system. The measurements were performed in
the frequency band near $800$ MHz. The experiment was set up
in the 5th floor of Building A of AT\&T's Middletown facility.
Three transmitters were set up in different parts of a wing
and the receiver moved through the wing.
A total of $200$ \emph{simultaneous channel sounding measurements for the
three transmitters were made}. 
The wideband path loss and the multipath delay profile 
of each transmitter were stored for each location.
The X \& Y coordinates of the floor map image, corresponding to the measurement
location, were saved, as well.

Fig.~\ref{fig:CenterTxData} and Fig.~\ref{fig:PatlabTxData} 
plot the wideband path loss of two transmitters as a heat map 
on the floor plan layout of the wing.
The X \& Y ticks in Fig.~\ref{fig:CenterTxData}-\ref{fig:StairTxData}
denote distances in meters.
In all these figures, the star and the circles show the transmitter
and measurement locations respectively. The height of the 
indoor transmitter 1, 2
and the receiver were $45$, $94$ and $47$ inches from the $5^{th}$ floor level.

Fig.~\ref{fig:CenterTxData} and~\ref{fig:PatlabTxData} show that
the path loss becomes higher as the receiver
moves away from the transmitter.
Since transmitter 2 is placed
in the central location of the wing, the mean path loss from 
transmitter 2 is lower than that from transmitter 1. Therefore, 
transmitter 2 will require less power than transmitter 1
to cover the whole wing.

The RMS delay spreads, averaged across all the measurement points
in the wing, were found to be $69$ ns and $72$ ns for transmitter $1$ and $2$ respectively.
We skip presenting the delay profiles of the 
measurement locations for brevity.

\subsection{Frequency domain channel sounding results}

\begin{figure}[t]
\centering
\includegraphics[scale=0.7]{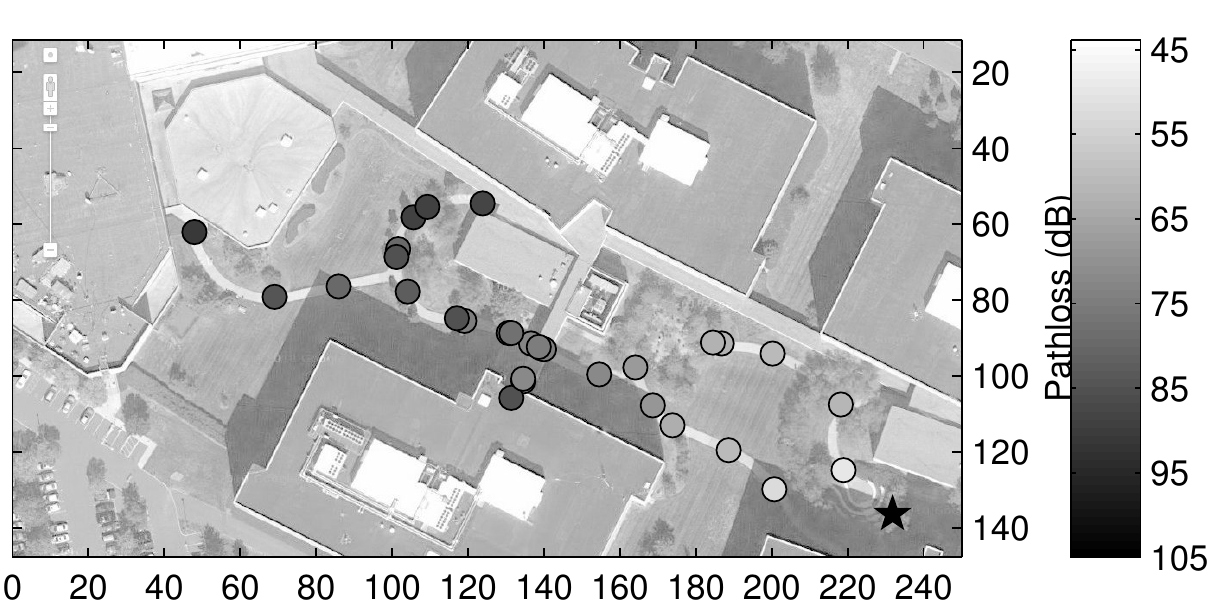}
\caption{Path loss for outdoor transmitter 2}
\centering
\label{fig:StairTxData}
\end{figure}


The outdoor channel measurements were obtained using
the frequency domain channel sounding system. The measurements were performed in
ten discrete steps of $2$ MHz and in the frequency band near $700$ MHz. 
The experiment was set up in the courtyard of Building A of AT\&T's Middletown facility.
Two transmitters were set up in two different corners of the
courtyard. 
In total, $50$ channel sounding measurements were taken \emph{simultaneously for
each transmitter} in different locations of this court yard. 
The height of the outdoor transmitter 1, 2 and the receiver
were 6, 12 \& 3 feet respectively from the ground level.
The ten narrowband path loss measurements
of each transmitter were stored for each location.
The GPS location~\cite{GPSsync} and the X \& Y coordinates in the satellite image view 
were saved, as well.

Fig.~\ref{fig:TripodTxData} and~\ref{fig:StairTxData} 
plot the mean wideband path loss of outdoor transmitter 1 and 2 as a heat map 
on the satellite image view of the courtyard. 
A comparison between Fig.~\ref{fig:CenterTxData} \&~\ref{fig:PatlabTxData} and
Fig.~\ref{fig:TripodTxData} \&~\ref{fig:StairTxData} reveals that the outdoor path losses
decrease less rapidly than the indoor ones. This happens because the outdoor 
signal does not get attenuated through any wall.

\begin{figure}[t]
\centering
\includegraphics[scale=0.45]{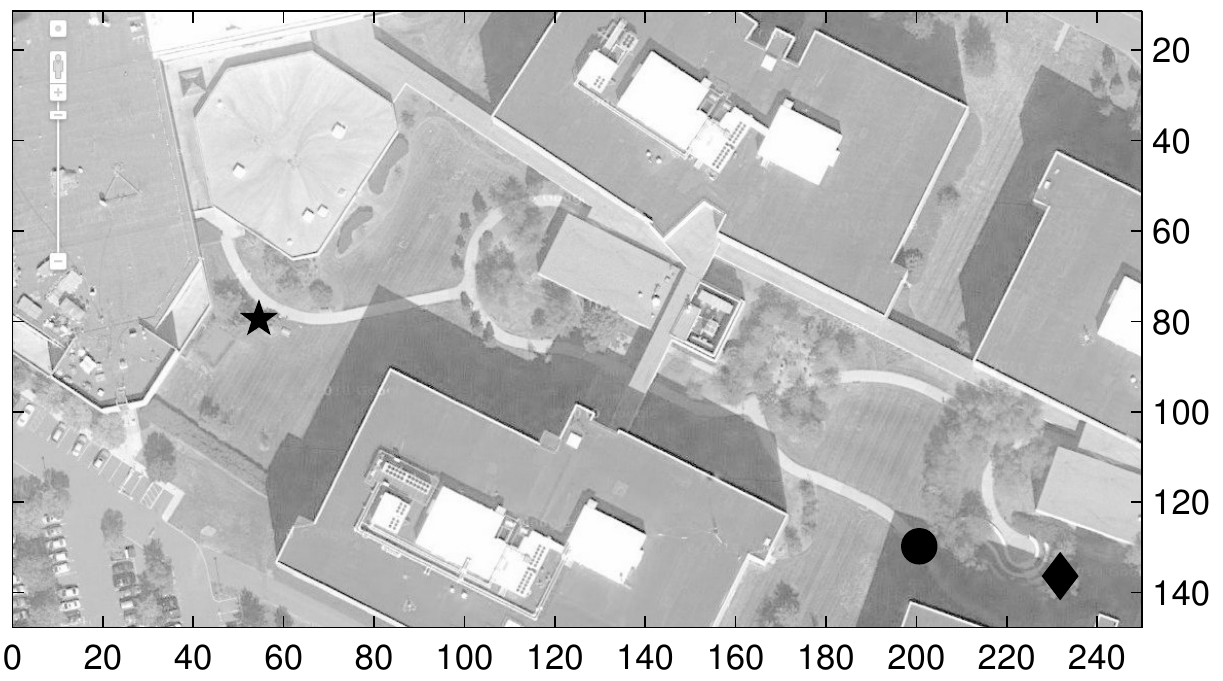}
\caption{Location of the transmitters and the measurement point}
\centering
\label{fig:FreqVariationLocation}
\end{figure}
\begin{figure}[t]
\centering
\includegraphics[scale=0.4]{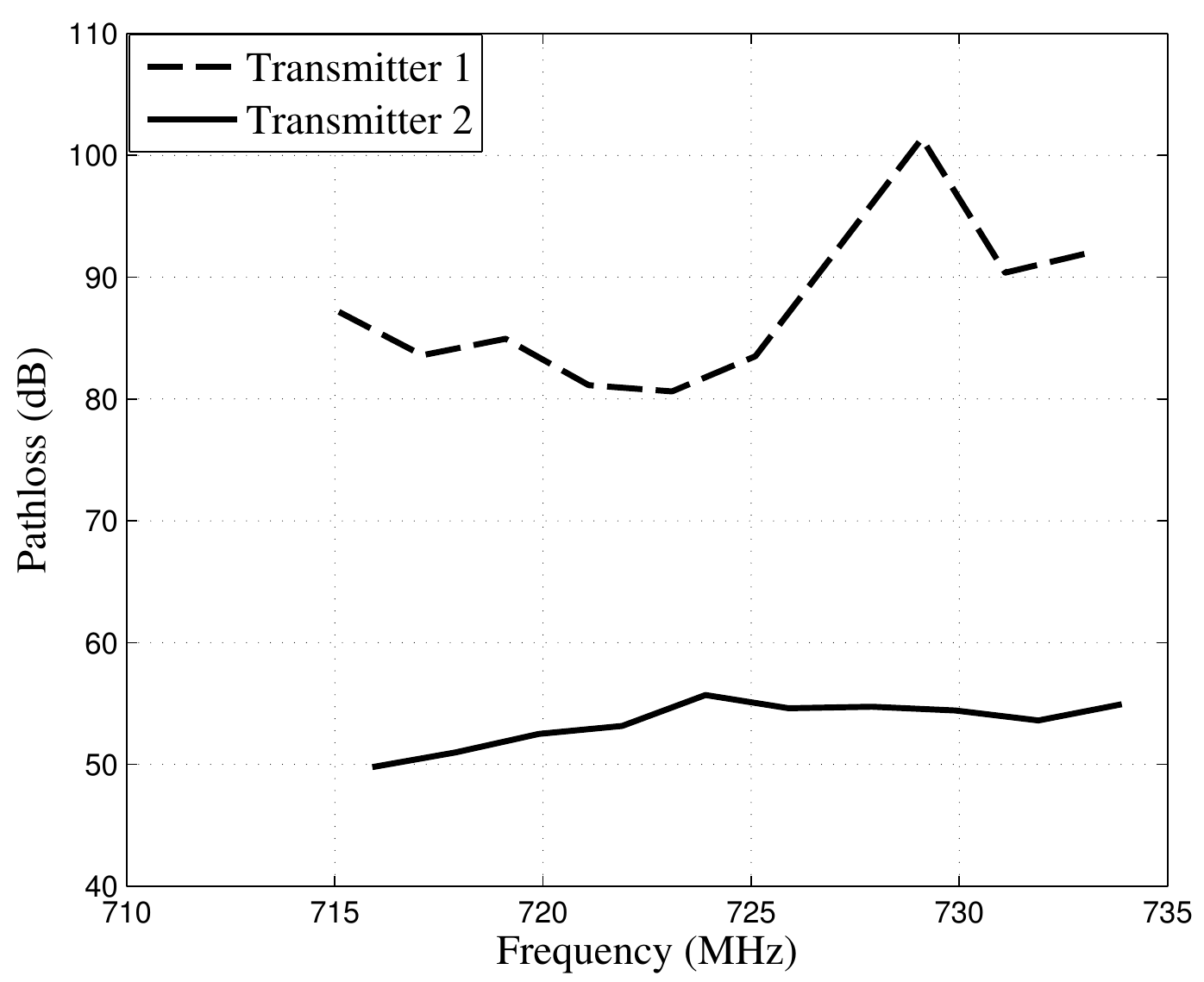}
\caption{Narrowband path loss of both transmitters}
\centering
\label{fig:FreqVariation}
\end{figure}

We now focus on variation in the narrowband path losses across the 
frequency band. The star, diamond and circle shapes in 
Fig.~\ref{fig:FreqVariationLocation} show the locations of outdoor transmitter $1$,
transmitter $2$ and the receiver respectively for a particular
measurement. Fig.~\ref{fig:FreqVariation} shows the path loss
spectrum of transmitter $1$ and $2$ at the receiver location. 
The path loss from transmitter $2$ varies only by $5$ dB in the $18$ MHz
band. This happens since the receiver is located very close to transmitter 2
and therefore, it does not experience much multipath from transmitter 2.
On the other hand, the receiver is located in a far
and non-line-of-sight location from transmitter 1.
Therefore, it experiences rich multipath from transmitter 1
due to the nearby buildings and foliage. Fig.~\ref{fig:FreqVariation} shows that
the path loss from transmitter 1 varies by $20$ dB across the frequency band.


\section{Conclusion}

We design and implement rapid wireless channel sounding systems,
using both sliding correlator and frequency domain approaches. 
Our design measures the 
channel propagation characteristics simultaneously from multiple transmitter locations.
Thus, the proposed design allows researchers to
quickly verify channel models with real data. 
It also assists engineers to compare the coverage
of multiple transmitter sites with a single run of measurements. 

Our achieved temporal resolution ($60$ ns) in the sliding
correlator method is limited by the data transfer rate
of the ethernet interface. The implementation of our algorithm directly on the
USRP FPGA will avoid this bottleneck and obtain a
temporal resolution of $10$ ns. This extension remains an area of future research.


For simplicity and time, we decided to design without explicit coordination 
between transmitters and receivers. Our future efforts will focus on more coordination 
through explicit communication to dynamically change 
 power, timing, frequency, and other aspects of the system.

\section{Acknowledgements}

We thank the members of the GNU Radio mailing list
for their continuous support throughout the project.

\bibliographystyle{IEEEbib}
\bibliography{BibICC2013}

\end{document}